\documentclass[12pt]{article}
\usepackage[utf8]{inputenc}
\usepackage[a4paper]{geometry}
\usepackage{graphicx}
\usepackage[textwidth=8em,textsize=small]{todonotes}
\usepackage{amsmath}
\usepackage{url}
\usepackage[figuresright]{rotating}
\usepackage{siunitx}
\usepackage{comment}
\usepackage{amssymb}
\usepackage{makecell}
\usepackage{booktabs}
\usepackage{caption}

\def\bSig\mathbf{\Sigma}

\newcommand{\myvec}[1]{ \mathbf{#1} }
\newcommand{\mymat}[1]{ \boldsymbol{#1} }

\newcommand{\mygvec}[1]{ \boldsymbol{#1} }

\usepackage{color,colortbl}

\definecolor{Gray}{gray}{0.5}

\title{Generalised Linear Models for Dependent Binary Outcomes
with Applications to Household Stratified Pandemic Influenza Data}
\author{Timothy Kinyanjui$^{1*}$ \& Thomas House$^1$}
\date{%
    $^1$Department of Mathematics, University of Manchester, Manchester, Oxford Steet, M13 9PL, UK\\%
    $^*$corresponding author: timothymuiruri.kinyanjui@manchester.ac.uk\\
}
\begin{document}

\maketitle

\begin{abstract}
Much traditional statistical modelling assumes that the outcome variables of
interest are independent of each other when conditioned on the explanatory
variables. This assumption is strongly violated in the case of infectious
diseases, particularly in close-contact settings such as households, where each
individual's probability of infection is strongly influenced by whether other
household members experience infection. On the other hand, general multi-type
transmission models of household epidemics quickly become unidentifiable from
data as the number of types increases. This has led to a situation where it
is has not been possible to draw consistent conclusions from household studies
of infectious diseases, for example in the event of an influenza pandemic.
Here, we present a generalised linear modelling framework for binary outcomes
in sub-units that can (i) capture the effects of non-independence arising from
a transmission process and (ii) adjust estimates of disease risk and severity
for differences in study population characteristics. This model allows for
computationally fast estimation, uncertainty quantification, covariate choice
and model selection. In application to real pandemic influenza household data,
we show that it is formally favoured over existing modelling approaches.
%We then do
%model selection using Akaike Information Criterion (AIC) and augment it by
%developing the Pseudo Wald's W statistical test for correlated outcomes. We
%demonstrate that accounting for group dependence using a disease transmission
%model coupled with a regression model improves the utility of the framework
%over standard generalised linear regression.
% Please include a maximum of seven keywords
%\keywords{Generalised linear regression, final size data, stochastic household models, Hypothesis test}
\end{abstract}

\section{Introduction}

\label{sec1}

Traditionally, epidemiological studies have focused on the identification of
individual risk factors for disease with the assumption that the cause of the
desired effect can be found at the individual level
\cite{Casado2014,Brown2015,Shi2015}. Here, populations are regarded as
collections of essentially independent individuals instead of as entities with
intrinsic properties that can be linked to a person's risk of developing
disease. The idea that population-level or more specific group-level factors
are important in understanding the distribution and acquisition of an
infectious disease has been well appreciated for a long time. A good example of
this is herd immunity i.e.\ the risk of contracting an infectious disease
depends in part on the level of immunity in the group to which they belong
\cite{Anderson1985a,Anderson1991,Rohani2008}. Herd immunity is therefore a
group property that is important in understanding the population level
transmission and the individual risk of infection and which is not captured by
group-level models such as multi-level analysis that still maintain independent
outcomes \cite{Snidjer2012}.

Models of infection and the associated non-independence in household models
were some of the earliest in mathematical epidemiology \cite{Bailey:1957}, and
were considered in particular theoretical depth in the influential paper by
\cite{Ball1997}.  While a `multi-type' version of this model, in which
individuals can have differing risks, can be fit to data using computationally
intensive methods such as Markov chain Monte Carlo \cite{Demiris:2005}, during
the 2009-10 influenza pandemic only a minority of studies of household
transmission made use of transmission models, with the majority using analysis
methods that did not account for non-independence of outcomes and hence the
transmissibility of influenza \cite{House2012}. On the other hand, in the
review and meta-analysis of \cite{Lau:2012}, the Forest plots for the
Secondary infection risks showed a lack of a consistent value in different
household studies; since these had a very different population, there is
potentially a lot of value to development of methods to make consistent
inferences through adjusting for such differences.

Regression is, in general, concerned with describing the relationship between a
response variable and a number of one or more independent variables usually
referred to as covariates. One of the classical uses of generalised linear
regression in epidemiology is in describing a binary outcome that is dependent
on a number of covariates. For example, a researcher might be interested in the
association between multiple independent covariates and the development of
disease, which is a binary outcome \cite{Hosmer2013,Cox1989}. In many cases,
participants in such a study will share an environment that can elevate their
probabilities of getting infected simply because of close proximity e.g.\ for
the case of respiratory infections in people who live in the same household or
are co-located in a shared environment. It has been shown that sharing living
arrangements can increase the likelihood of an infection spreading to other
members of a household \cite{Black2014,Tsang2017,Kinyanjui2018} and a recent
individual based modelling study \cite{Kombe2018} has found that household
transmission structuring is important in explaining co-existence of RSV group A
and B and their differential transmissibility. In such instances, standard
generalised linear regression models that assume independence between outcomes,
would fall short of explaining the observations through group membership. The
question of dependence between observations \emph{via group membership} has
been addressed using multi-level, also known as Hierarchical linear models
\cite{Bryk1992}.  Multi-level models are useful for providing an improved
estimate of effect within individual units in a nested structure e.g.\
developing an improved regression model for an individual's risk of disease
given the fact that they share a household/environment with other members who
might be exposed to the same risks such as environmental exposures modelled
using a random effects approach. However, when considering an infectious
disease outcome e.g.\ whether a person becomes infected in a household or not,
a more appropriate construction would include a disease process model in which
one individual's outcome (infection) is another individual's exposure (through
contact with an infective).

In the rest of the work, we will present a unified approach to household
modelling by fusing a stochastic epidemic model and a generalised linear
regression model to describe the data from a household study conducted in Spain
during the 2009-2010 influenza pandemic \cite{Casado2014}. We assess model
performance against standard generalised linear regression and undifferentiated
household models, and argue that this approach could be usefully applied in the
case of a future pandemic or other outbreak in which multiple household studies
are performed.

\section{Methods}
\label{sec2}

In our methodological development we will follow convention and refer to
infection in households, despite the more general nature of the approach, which
can apply to any contagious process in groups. We start by presenting the two
components of our model.  The first component comprises a stochastic epidemic
model incorporating additional biologically relevant information on the nature
of outcome dependencies and the second component comprises a linear regression
model for binary outcomes. In the next section, we will first motivate the
statistical methodology and then describe the application of the method to
Spanish influenza household data collected by \cite{Casado2014}.

\subsection{Statistical analysis}
\subsubsection{General stochastic contagion model}\label{gsm}

\cite{Ball1986} proposed a stochastic epidemic model
that allows for a flexible analysis of final size household infection data. The
model we consider here is an extension that allows for multiple sources of infections,
in addition to
variable length of infectious period, heterogeneous contact rates reflecting
variable susceptibility and infectiousness as well as mixing behaviours. For completeness, we describe the model below
but for a full description, we refer the reader to \cite{Addy1991}.

Let $P^{(N_1,N_2,\ldots ,N_m)}_{(\omega_1,\omega_2,\ldots ,\omega_m)}$ be the
final size probabilities i.e. the probability that $(\omega_1,\omega_2,\ldots
,\omega_m)$ susceptibles get infected in a household with initial susceptibles
being $N$ such that $N = \sum_{i=1}^{m}N_i$. As discussed above, we
will follow the literature in calling the group / sub-population
a household, and each person within the
household a type. We can write the final size probabilities succinctly as
$P^{\myvec{N}}_{\mymat{\omega}}$ such that $\myvec{N}=(N_1,N_2,\ldots ,N_m)$
and $\mymat{\omega}=(\omega_1,\omega_2,\ldots ,\omega_m)$. Then for each
household, the final size probabilities can be given by 
\begin{equation}
\frac{%
\sum_{\omega_1=0}^{j_1}\sum_{\omega_2=0}^{j_2}\cdots \sum_{\omega_m=0}^{j_m}\binom{j_1}{w_1}\binom{j_2}{w_2}\cdots\binom{j_m}{w_m}P^{\myvec{N}}_{\mymat{\omega}}}%
{\prod_{k=1}^{m}{\rm e}^{-\Lambda_{k}^{(N_k-j_k)}}
\phi_k\left(\sum_{i=1}^m (N_i-j_i)\lambda_{ik}\right)^{\omega_k+a_k}} = 1 \text{ .}
\label{addys}
\end{equation}
Note that each $\phi_k$ term in the product in the denominator is the
probability that the $N_i-j_i$ susceptibles of each type $i$ avoid infection
from $\omega_k+a_k$ infectives in group $k$. The product then becomes the
probability that all indexed susceptibles avoid all the indexed infectives for the entire
duration of their respective infectious period. It is worth highlighting at
this point that $\lambda_{ik}$ governs within-population disease transmission
and is the rate at which a susceptible of type $i$ has contact with an
infective of type $k$ and hence it is an $m\times m$ matrix. It can therefore
be structured to model different assumptions e.g. a model of variable
susceptibility and fixed infectivity would have $\lambda_{ik}=\lambda_{i}$ so
that it only depends on the group of susceptibles. Note that to determine the
final size probabilities $P^{\myvec{N}}_{\mymat{\omega}}$, then we need to
solve the resulting system of linear equations from Eqn. \ref{addys}.

%The phenomenon that the probability of an individual being infected, e.g.\ in
%the case of an infectious disease, depends in part on the prevalence of the
%infection in the group to which they belong, in our case a household, and also
%may be related to other person or group specific characteristics/covariates.
%During the most recent pandemic, however, the overwhelming majority of studies
%did not account for the effect of transmission, and those that did typically
%included relatively few covariates \cite{House2012}.
%As we have a stochastic model developed in
%section \ref{gsm}, the novelty of our work lies in combining it with a 
%generalised linear
%regression framework hence creating a unified model.

Let us consider Eqn.~\ref{addys} and express it more succinctly as 
\begin{equation}
    \mathcal{B}(\Lambda_k,\lambda_{ik},\theta,\alpha)\myvec{P}=\mathbf{1}.
    \label{succ}
\end{equation}
Here $\myvec{P}$ is a vector formed of the $P^{\myvec{N}}_{\mymat{\omega}}$, in
our case by lexicographical ordering, and $\mathcal{B}$ is the matrix implied by
Eqn~\eqref{addys} under this ordering.  The parameter $\Lambda$ represents the
global or between household probability of transmission. $\lambda$ represents
the within household transmission which is, as is often done \cite{Cauchemez2004,Cauchemez2011},
scaled with household size as $\lambda\propto (N-1)^{-\alpha}$, with
$\alpha$ representing the different ways that mixing behaviour can change with
household size. If $\alpha=0$, then every pair of individuals make contacts
capable of spreading the infection at the same rate and if $\alpha>0$, then a
larger household reduces the rate of transmission. The stochastic model
presented by Eqn. \ref{addys} allows for any distribution of the length of the
infectious period provided that its Laplace transform, $\phi$, can be
specified.  We model the length of infectious period using the Gamma
distribution with variance $\theta$ and unit mean (since the final size of
an epidemic is insensitive to the choice of mean) giving
\begin{equation}
\phi_k(s) = \left( 1 + \theta s\right)^{-1/\theta} \text{ ,}\quad \forall k\text{ .}
\label{gamm}
\end{equation}
Practically, we can then solve Eqn.~\eqref{succ} numerically using standard
linear algebra techniques. In terms of model fitting, however, there are many
parameters involved (most notably the $\lambda_{ik}$) and as such we need a
strategy to reduce the number of these to achieve identifiability.

\subsubsection{Regression model}

\label{regsec}

Now suppose that we have an individual $i$ at a fairly constant risk of
infection $\Lambda_i$ over a unit time period. The probability of this
individual's escaping infection is given by $\exp{(-\Lambda_{i})}$. Rather than
estimating the probability of escaping the infection, we are will model the
probability of being infected, which is $\pi_{i}=1-\exp{(-\Lambda_{i})}$.
It follows that 
\begin{equation}
-\log(1-\pi_{i})=\Lambda_{i}\text{ .}
\label{loglog}
\end{equation}
If we want to model a rate such as $\Lambda \in [0,\infty)$ in terms of a linear predictor of covariates $\myvec{X}_i$ of $i$, then the natural choice would be a log-linear model
\begin{equation}
\log(\Lambda_{i})=\mygvec{\beta}\cdot \myvec{X}_{i}\text{ .}
\end{equation}
Eqn. \ref{loglog} then becomes 
\begin{equation}
f(\pi_{i}) = \log(-\log(1-\pi_{i}))=\mygvec{\beta}\cdot \myvec{X}_{i}.
\label{loglogmodel}
\end{equation}
The function $f:[0,1]\rightarrow\mathbb{R}$ therefore functions as a sigmoidal
link function, and \eqref{loglogmodel} is usually called the complementary
log-log regression model~\cite{McCullagh:1980}.  The link function $f$ is preferable to the more
commonly used logistic and probit functions when a rate is involved due to
the interpretability of regression coefficients. Maximum likelihood estimation
for data $D=(y_1,\ldots,y_n)$, where $y_i=1$ if individual $i$ experiences
infection and $y_i=0$ otherwise, is possible by writing
\begin{equation}
\pi_i = 1 - \exp(-\exp(\mygvec{\beta}\cdot\myvec{X}_i)) \text{ ,} \qquad
L(D|\mygvec{\beta}) = \prod_{i=1}^{n} \pi_i^{y_i}(1-\pi_i)^{(1-y_i)} \text{ ,}
\end{equation}
and then finding a maximum of the likelihood function $L$.  Such an approach
has the benefit of the incorporation of various person--specific characteristics
which are thought to influence the acquisition of the infection (see section
\ref{datadescription} for a description of the data) but assumes, potentially
incorrectly that outcomes are independent.

\subsubsection{Unified models}

We consider the following four scenarios of incorporating regression model
within the generalised stochastic model in Eqn \ref{succ}.

In the first scenario, which we denote as {\it{HH-$\lambda$}}, we let the parameter
that governs the within-household disease transmission, $\lambda_{ik}$, depend
on the susceptible person such that $\log(\lambda_{ik})=\mygvec{\beta}\cdot
\myvec{X}_{i}$.

In the second scenario which we denote as, {\it{HH-$\Lambda$}}, we let the
parameter that governs between-household disease transmission, $\Lambda_{i}$,
depend on the covariates such that
$\log(\Lambda_{i})=\mygvec{\beta}\cdot\myvec{X}_{i}$.

In the third scenario, which we denote as {\it{HH-Both}}, we let both the
within and between household transmission parameters depend on the co-variates
simultaneously i.e.\ a combination of the first and second scenarios.

In the forth scenario, which we denote as {\it{HH-Null}}, we let the within and between household
transmission parameters vary freely without dependence on any covariates i.e. they
are not related in anyway to the regression model.

The final scenario, denoted as {\it{Reg}}, is the regression model described in
\S{}\ref{regsec}. In this model, observations are assumed independent and
therefore within household relationship cannot be accounted for. Because this
has been the standard model in use by researchers in the field of
epidemiology, we adopt it as our baseline against which we judge the
performance of the other methods, and the objective of this work is to
improve its performance by allowing the probability of an individual getting
infected depend on the within and between household transmission probability.

We note that these possibilities are intended to be indicative rather than
exhaustive, and that other possibilities such as splitting covariates into
those expected to influence susceptibility versus transmissibility, and within-
versus between-household transmission, is likely to be the most pragmatic modelling approach
in applications of our methodology. 

\subsubsection{Likelihood calculation and model fitting}

Solving for $\myvec{P}$ in Eqn.~\eqref{succ} gives us the probability that a
household is in a certain final size configuration. To make it clearer, we will
give an example here.  Suppose we have a household with two initially
susceptible individuals. Then solving Eqn.~\eqref{succ} gives us the
probabilities associated with all the possible infection configurations i.e.\
$P^{(1,1)}_{(0,0)},P^{(1,1)}_{(1,0)},P^{(1,1)}_{(0,1)}$ and
$P_{(1,1)}^{(1,1)}$. We know the final size configurations for household $i$
from data, which we denote as $D_i=(y_{i,1},y_{i,2},\ldots,y_{i,k_i})$, where
$k_i$ is the household size. Each of the $y_{i,j}$ is an indicator variable
taking the value $1$ if a individual $j$ in household $i$ is infected and $0$
otherwise. For household $i$, the likelihood of observing the data $D_i$ is
given by ${\ell}_i(D_i\mid
\mygvec{\vartheta})={P}_{D_i}^{\mathbf{1}}(\mygvec{\vartheta})$ where
$\mygvec{\vartheta}$ are the model parameters that need to be estimated and
$\mathbf{1}$ is a length-$k_i$ vector of ones. The total likelihood therefore
becomes
\begin{equation}
   L(D\mid \mygvec{\vartheta}) = \prod_{i=1}^m{\ell}_i(D_i\mid \mygvec{\vartheta})\; .
   \label{likelihood}
\end{equation}
Note that in general
$\mygvec{\vartheta}\subseteq(\Lambda,\lambda,\theta,\alpha,\mygvec{\beta})$,
where the vector of regression coefficients is $\mygvec{\beta} =
(b_i)_{i=0}^{14}$, and the exact parameters estimated for each model are
shown in Table~\ref{table1}.

We estimate the model parameters numerically by fitting the model to data
$D=(D_1,\ldots,D_m)$ using maximum-likelihood methods based on the likelihood
as shown in Eqn \ref{likelihood}. The negative log-likelihood function was used
as the objective function in a numerical  minimization
routine using Quasi-Newton methods. To calculate the $95\%$ CIs of the fitted parameters,
we computed the central finite difference approximation to the Hessian of the negative
log-likelihood estimates to generate an asymptotic covariance matrix and then
used a normal approximation \cite{Dennis1996,Cox1974} to estimate the confidence region.

\subsubsection{Hypothesis testing}

We will wish to test for statistical significance of regression co-efficients
$\mygvec{\beta} =
(b_i)_{i=0}^{14}$, which is possible using a Pseudo-Wald's W test. Following
the discussion in
the previous section, we note that
$\mygvec{\vartheta}=(\Lambda,\lambda,\theta,\alpha,\mygvec{\beta})$ is a row
vector that denotes the most general set of parameters of the epidemic
model
that need to be estimated. To develop the test, we take the approach introduced
by Ball and Shaw \cite{Shaw2016}. Expressly, we want to test the models with
$b_i\neq0$ for $i\in\mathcal{I}\subseteq\{0,1,\ldots,14\}$ against the
null model in which $b_i=0$ for this set of regression coefficients. The test
can be written as follows:
\begin{equation*}
\begin{aligned}
H_0&~:~~b_i=0~~~~~i \in\mathcal{I} \; ,\\
H_1&~:~~b_i\neq0~~~~~i \in\mathcal{I} \; .
\end{aligned}
\end{equation*}
Suppose we want, for example, to test the hypothesis that all regression
parameters are zero. Then in terms of $\mygvec{\vartheta}$,
\begin{equation}\label{h1}
\begin{aligned}
H_0&~:~~\vartheta_5=\vartheta_6=...=\vartheta_{n_{max}}=0 \; ,\\
H_1&~:~~\vartheta_j\neq0~~~~~\text{for }~~ j\geq 5 \; .
\end{aligned}
\end{equation}
In general, let $\mygvec{h}(\mygvec{\vartheta})$ be a vector of length
$|\mathcal{I}|$ such
that $h_i(\mygvec{\vartheta})=\vartheta_{\mathcal{I}_i}$, where $\mathcal{I}_i$
is the $i$-th element of $\mathcal{I}$.
Then, the hypotheses can be re-written as
\begin{equation}\label{h2}
\begin{aligned}
H_0&~:~~\mygvec{h}(\mygvec{\vartheta})=\mygvec{0}\\
H_1&~:~~h_i(\mygvec{\vartheta})\neq0~~~~~\text{for }~~ i\in
\{1,2,\ldots,|\mathcal{I}|\} .
\end{aligned}
\end{equation}
Let $\mygvec{\hat{\vartheta}}$ denote the unrestricted maximum pseudolikelihood
estimator under $H_1$ and $\mygvec{H_\vartheta}$ be the
matrix with elements given by
$(H_{\mygvec{\vartheta}})_{ij} = \partial h_j / \partial
\vartheta_i$ so that, for the case of all regression parameters being zero
considered in Eqn.~\eqref{h1} above,
\begin{equation*}
\mygvec{H_\vartheta}=
\begin{bmatrix}
0 & 0 & 0 & \dots & 0\\
0 & 0 & 0 & \dots & 0\\
0 & 0 & 0 & \dots & 0\\
0 & 0 & 0 & \dots & 0\\
1 & 0 & 0 & \dots & 0\\
0 & 1 & 0 & \dots & 0\\
\vdots & \vdots & \vdots & \ddots & 0\\
0 & 0 & 0 & \dots & 1\\
\end{bmatrix}
\end{equation*}
The first four rows of $\mygvec{H_\vartheta}$ are a row of zeros due to
$(\vartheta_1,\vartheta_2,\vartheta_3,\vartheta_4)$ not appearing in our
constraint vector $\mygvec{h}\left(\mygvec{\vartheta}\right)$. The
Pseudo-Wald's W test
assumes that under the null hypothesis,
$\mygvec{h}\left(\mygvec{\vartheta}\right)=\mygvec{0}$ and therefore if $H_0$
is true,
we expect that
$\mygvec{h}\hat{\left(\mygvec{\vartheta}\right)}\approx\mygvec{0}$. From
Taylor's theorem, we can see that
$\mygvec{h}\hat{\left(\mygvec{\vartheta}\right)}\approx
\mygvec{h}\left(\mygvec{\vartheta}\right)+\mygvec{H_{\mygvec{\vartheta}}}(\mygvec{\hat{\vartheta}}-\mygvec{\vartheta})$.
Ball and Shaw \cite{Shaw2016} have shown that 
\begin{equation}\label{h3}
m\mygvec{h}\left(\hat{\mygvec{\vartheta}}^{(v)}\right)^{T}\left(\mygvec{H_{\vartheta}}^{T} \mygvec{I_{\vartheta}}^{-1} \mygvec{\Sigma_{\vartheta}} \mygvec{I_{\vartheta}}^{-1} \mygvec{H_{\vartheta}} \right)^{-1} \mygvec{h}\left(\hat{\mygvec{\vartheta}}^{(v)}\right)\xrightarrow{\mathcal{D}} \mathcal{\chi}_{|\mathcal{I}|}^{2}
\; ,
\end{equation}
where $m$ is the total number of households, $\mygvec{I_{\vartheta}}$ is the
Fisher information matrix with respect to $\mygvec{\vartheta}$ with components
$I_{ij}\left(D\mid\mygvec{\vartheta}\right)=-(\partial^2/\partial{\vartheta_i}\partial{\vartheta_j})L\left(D\mid
\mygvec{\vartheta}\right)$ and $\mygvec{\Sigma_\vartheta}$ is the covariance
matrix. The hypothesis test can therefore be carried out from Eqn.~\eqref{h3}
as the sampling distribution of the test statistic is a chi-squared
distribution when the null hypothesis is true. We will use this test to
calculate p-values for each regression coefficient -- while recognising the
criticisms that can be made of such an approach \cite{Colquhoun:2014} -- to
demonstrate the consistency of household regression with standard statistical
practice.

\subsubsection{Simulation strategy}

To explore the performance of the approach proposed here, we compare the
performance of the four model scenarios to the baseline which is the standard
complementary log-log regression model.  Overall model selection is performed
using the Akaike Information Criterion (AIC) \cite{Akaike:1974}, which allows
selection of models that do not meet the assumptions of the hypothesis tests
above while penalising excess complexity.

\subsection{Application to influenza data}

To demonstrate the applicability of the methods developed in the previous
sections, we use household influenza data to estimate the model parameters as
well as assess how well the model performs against the baseline.

\subsubsection{Description of the data}\label{datadescription}

The study was conducted in Navarra, Spain during the A(H1N1)pdm09 influenza season between 2009-2011. The primary surveillance network, comprising of physicians
and paediatricians took nasopharyngeal and pharyngeal swabs of all patients
diagnosed with influenza-like illness (ILI) whose symptoms had begun within the
previous 5 days. A public health nurse telephoned the households of each index
case and conducted a structured interview. The questionnaire administered
during the telephone interview asked detailed information about the index case,
socio-demographic data of other members of the household and the dates of
symptoms onset of other household contacts. Secondary household cases were
susceptible household contacts who had ILI within 7 days from the onset of
symptoms in the index case. In the study, occurrence of ILI in household
contacts was assessed using multivariate logistic regression analysis adjusted
for a number of person specific co-variates and their measure of association
was done using odds ratio. The following co-variates were used in the analysis;
age of contacts, gender, major chronic conditions, vaccination status, sharing
a bedroom with index case, number of household members, rural or urban
municipality or residence, age of the index case and influenza season. For more
information on the study, we refer the reader to \cite{Casado2014}.

We chose this dataset partly because it is publicly and fully available
without restriction, allowing for reproducibility of our results. There are two
minor limitations of this data, however. The first is that he households in the
data were selected based on the availability of an index case in the household.
This implies that the household sampling is not random and we are uncertain as
the extent to which this biases the results for the general population, however
the results will hold for the population of households with one index case.
The second is that while household membership is present in the data, there is
some grouping for anonymity and so a small amount of imputation needs to be
carried out.

\section{Results}\label{sec3}

\begin{figure}[!h]
\begin{center}
    \includegraphics[width=1\textwidth]{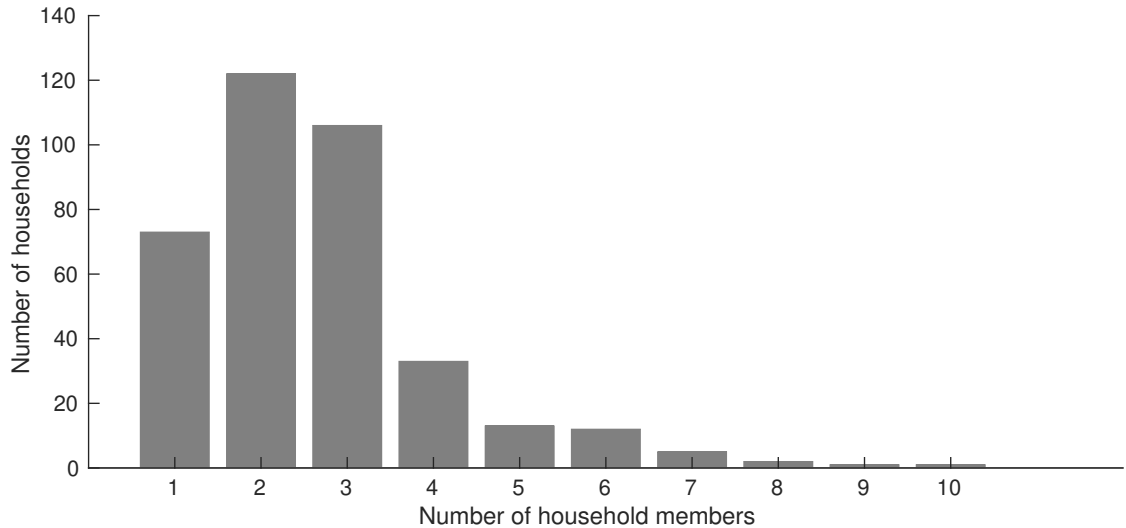}
\end{center}
\caption{Figure shows the final size data (grey bars) as frequency distribution of total number of cases stratified by household size.}
\label{hist}
\end{figure}

For this application, the household is considered as a sub-population and every
individual within the household is considered a type. The data has a total of
368 households with Figure \ref{hist} showing a histogram of the distribution
of household sizes. Households consisting of one, two or three members dominate
with a decreasing number of households with larger occupancy.

\begin{figure}[!hp]
\begin{center}
    \includegraphics[width=1\textwidth]{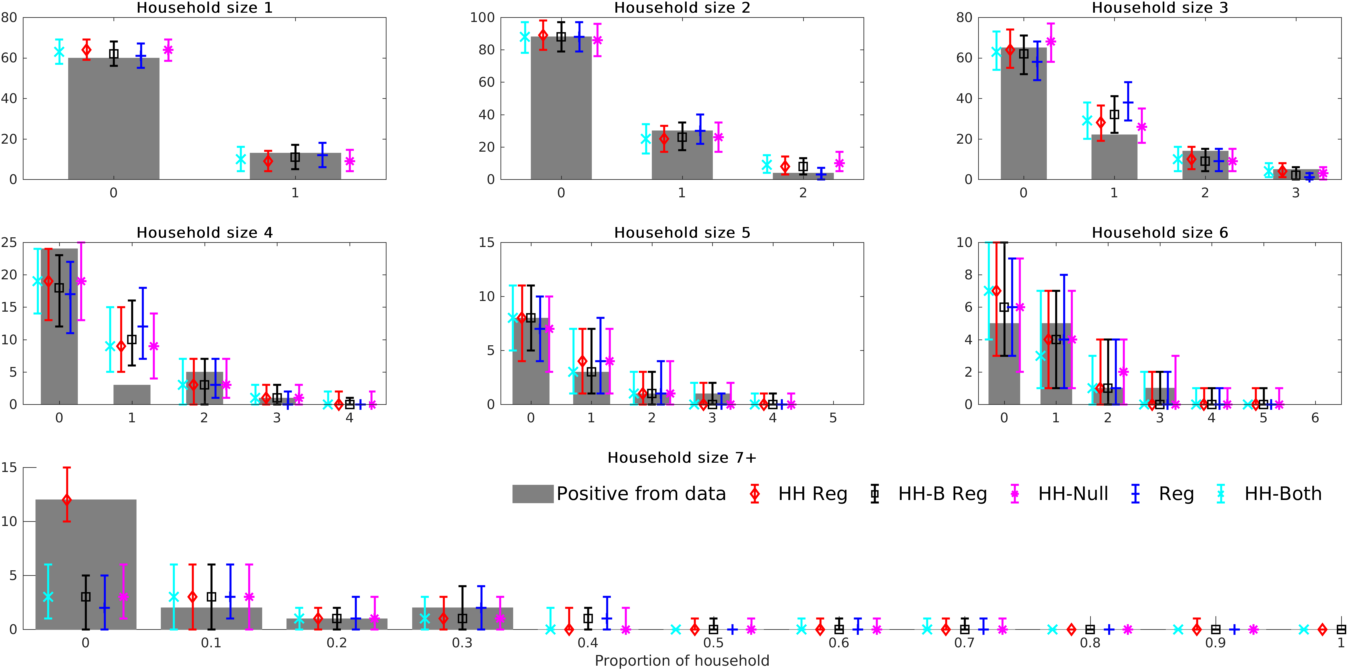}
\end{center}
\caption{Figure shows the final size data (grey bars) as frequency distribution of total number of cases stratified by household size.}
\label{figure1}
\end{figure}

\begin{figure}[!hp]
\begin{center}
    \includegraphics[width=1\textwidth]{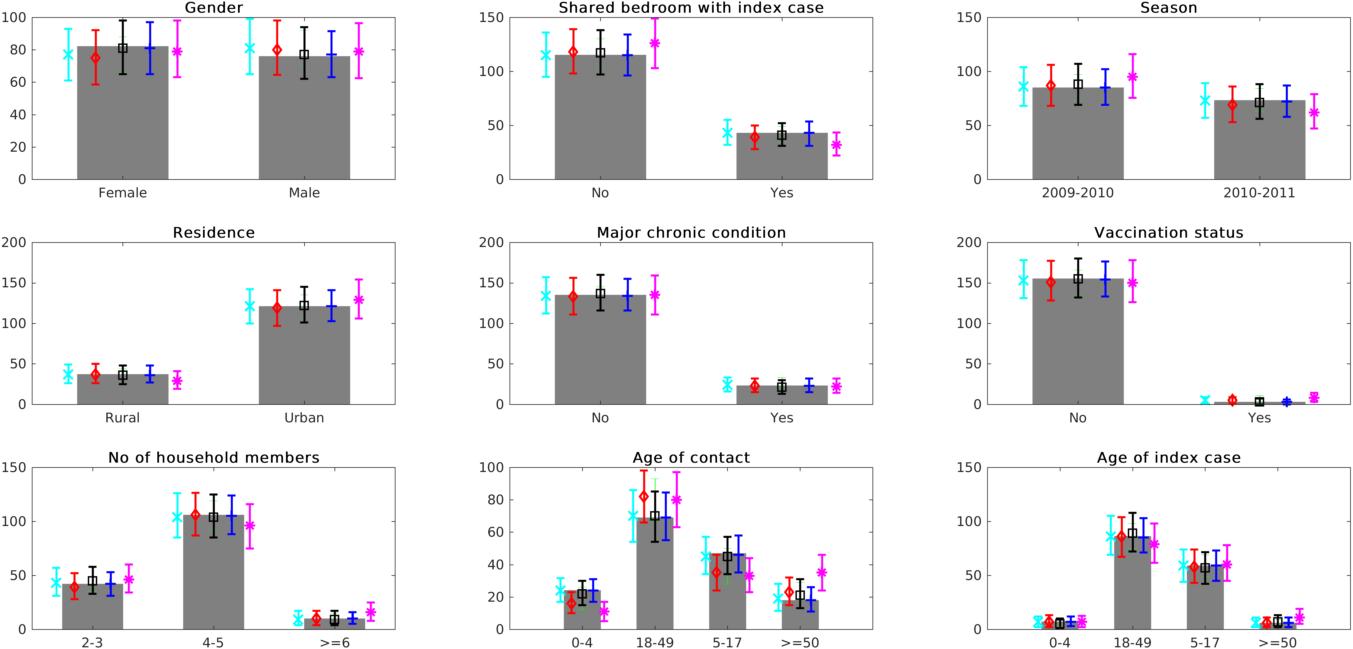}
\end{center}
\caption{Figure shows the final size data (grey bars) as frequency distribution of total number of cases stratified by co-variates.}
\label{figure2}
\end{figure}

Figures \ref{figure1} and \ref{figure2} shows how well the models describe the
(marginal distributions of) the observed data. The grey bars in each figure
represent the number of observed cases in the influenza data while the error
bars represent the final sizes predicted by the model for all the modelling
assumptions (see legend at bottom axes of Figure \ref{figure1}). Figure
\ref{figure1} shows the output by household size while Figure \ref{figure2}
shows the results by co-variates. We have further stratified the model fit in
Fig.~\ref{figure2} by households size (see Supplementary Figs.\ S1 to S7). The
results reveal that the models do fit well to the marginal household size
distributions for all the co-variates.

Table \ref{table1} shows the estimated model parameters and their $95\%$
confidence intervals for all the five modelling assumptions. The values to the
right of the CI intervals are the individual covariate p-values from the Wald
test. From the table, the co-variate whose p-value indicates significance
across all the models is $b_1$, which is household contacts in children under 5
years. It is also clear that gender ($b_{12}$) is not statistically significant
across all the models. The rest of the parameters have significance that is
dependent on the model selected.  From the AIC, which is a measure of the
relative goodness of fit, the best fitting model is HH-Both. This model has
$b_1$ and $b_{14}$ (participant having received flu vaccine) being
statistically significant. The bottom row shows the p-values from the
hypothesis test of the HH-Null model $(H_0)$ against the other three nested
models that have the regression model attached to the transmission parameters.
Observations of the table show that the unified models
(HH-$\lambda$, HH-$\Lambda$, HH-Both) have advantages over the HH-Null model
leading to such small p-values being observed. We can therefore reject the Null
model which assumes that within- ($\lambda$) and between- ($\Lambda$) household
transmission probabilities do not depend on individual or household level
co-variates.

\section{Discussion}

When a certain outcome is dependent on a number of factors that can be measured
or imputed, this is a
problem that properly renders itself to regression analysis. Traditional
regression analysis, however, assumes independence between observations and this is
usually not the case especially in infectious disease transmission where
sharing a common environment with an infected person can elevate ones risk of
contracting the infection. For example, a recent modelling study
\cite{Kombe2018} using time resolved RSV infection data for RSV found evidence
that there might be some niche (household) separation for the two RSV groups
explaining how the two groups are able to co-exist together within the same
epidemic. This is of course coupled with other factors such as weak cross
immunity, differential susceptibility and household composition.

To account for the dependence between observations, multi-level models have
been proposed and successfully used in various applications
\cite{Diez-Roux2000}. However, multi-level models are usually unable to
capture the feedback relationships that sometimes exist between predictors and
outcomes. 

In this paper, we explore a dependent-outcome generalised linear model that
aims to better detect re-infections probabilities than the standard  single
level linear models typically carried out. This is done by fusing a disease
transmission stochastic model that acts at the group level and then linking
disease transmission potential to individual characteristics using a regression
model. The stochastic contagion model presented offers a flexible statistical
tool for modelling infectious diseases for which the final size data is
available. It has an advantage that it makes use of a variable infectious
period whereas most previous work have incorporated a constant period or
adopted an assumption of an exponential distribution. The infectious period is
generalised in our current model so long as the Laplace transform can be
specified. In our case, we posited a gamma distribution as a good approximation
and evidence from previous simulation studies support our choice
\cite{Addy1991,House2012,Rohani2008}.

As we are introducing a method rather than testing specific biological
hypotheses, we proposed various indicative scenarios for linking co-variates to rates in section 2.1.3. We incorporated the regression on the within household transmission (HH-$\lambda$), between household transmission (HH-$\Lambda$), on both between and within households transmission (HH-Both) or we did not incorporate regression (HH-Null). In all scenarios, the unified transmission-dynamic models performed better, as measured by their p-values, see bottom row \ref{table1},
compared to the Null model (HH-Null).

As explained by the AIC, the model that best describes the data is HH-Both which governs both the within and between household disease transmission, an observation which is also supported by the statistical test. The next best model is HH-$\Lambda$. A potential explanation why both of these perform better than HH-$\lambda$ is that the covariates in the data are best associated with explaining between household transmission potential.

Our study is not without limitations. While the modelling framework is flexible enough to accommodate
a risk of within-household infection that is dependent on both the susceptibility profile of the
contacted person and the infectivity profile of the infectious contact, due to
limited data for estimating the model parameters, we assumed
variable susceptibility and fixed infectivity so that we have $n\times 1$
rather than $n\times n$ within household transmission rate $\lambda$. A future
extension would be to have both variable susceptibility and infectivity.

While AIC is perhaps one of the more popular tools for model selection, and included in this work for completeness, it would seem inadequate by itself given that it relies on having independent data or else its asymptotic properties breakdown. We have therefore augumented AIC with the Wald's W test for model selection and this seems, so far, to be the best model selection method for household epidemic data with correlated outcomes. However, standard methods for model selection could be employed if it can be shown that the dependence is weak, particularly if the number of households, $m$, in the data is large (note that dependence is of order $1/m$ \cite{Shaw2016}). In our case, this can not be justified.

In conclusion, our analysis shows that accounting for group dependence using a
disease transmission model coupled with a regression model improves the
predictive utility of the framework over the standard linear model. As such we
hope to have aided the design of study analysis plans where the assumption of
independence between observations does not hold and where dependence is
mechanistically linked, e.g.\ through close contacts, justifying a stochastic
contagion model.

\clearpage

\begin{sidewaystable}[p]
    \caption{Estimated model parameters, their (95\% CIs) and their p-values.}
    {\scriptsize \begin{tabular}{cccccc}
        \toprule
        $\mygvec{\vartheta}$ & HH-$\lambda$ &  HH-$\Lambda$ & HH-Both & HH-Null & Reg\\
        \toprule
        $\Lambda$ & 0.948(0.925,~0.971) & - & - & 0.948(0.912~0.983) & - \\
        $\lambda$ & - & 0.059(0.026,~0.091) & - & 0.096(0.058~0.134) & -\\
        $\theta$ & 2.49(2.1,~2.89) & 26.24(25.833,~26.643) & 7.502(6.251,~8.754) & 7.28(7.251~7.301) & -\\
        $\alpha$ & 1.136(0.815,~1.456) & 1.278(0.912,~1.644) & 1.408(1.104,~1.712) & 1.088(0.82~1.356) & - \\
        $b_0$ & $-$5.196($-$5.487,~$-$4.905)~$<$0.00001 & $-$4.34($-$4.633,~$-$4.048)~0.035 & $-$4.247($-$4.618,~$-$3.876)0.12 & - & $-$3.258($-$4.52,~$-$2.135)$<$0.00001\\
        $b_1$ & 2.837(2.39,~3.284)~$<$0.00001 & 2.221(1.907,~2.536)~$<$0.00001 & 2.387(2.101,~2.672)0.02 & - & 1.723(1.07,~2.4)$<$0.00001\\
        $b_2$ & 1.03(0.703,~1.357)~$<$0.00001 & 0.647(0.368,~0.927)~0.38 & 0.471(0.19,~0.751)0.75 & - & 0.44($-$0.102,~1.029)0.13\\
        $b_3$ & 1.619(1.253,~1.984)~$<$0.00001 & 1.585(1.251,~1.919)~0.051 & 1.638(1.334,~1.942)0.26 & - & 1.149(0.568,~1.773)0.0002\\
        $b_4$ & 0.987(0.622,~1.353)~$<$0.00001 & 0.018($-$0.278,~0.314)~0.66 & 0.485(0.217,~0.752)0.38 & - & 0.601($-0.56$,~1.797)0.31\\
        $b_5$ & 0.928(0.598,~1.258)~$<$0.00001 & 0.506(0.19,~0.822)~0.71 & 0.317(0.003,~0.631)0.87 & - & 0.421($-$0.374,~1.4)0.34\\
        $b_6$ & 0.974(0.636,~1.311)~$<$0.00001 & 0.454(0.128,~0.78)~0.48 & 0.474(0.161,~0.786)0.78 & - & 0.532($-$0.306,~1.546)0.25\\
        $b_7$ & 1.066(0.596,~1.535)~$<$0.00001 & 0.501(0.085,~0.917)~0.079 & 0.606(0.27,~0.943)0.52 & - & 0.433($-$0.061,~0.885)0.071\\
        $b_8$ & 0.926(0.598,~1.255)~$<$0.00001 & 1.359(0.983,~1.735)~0.028 & 0.882(0.633,~1.132)0.53 & - & 0.825(0.13,~1.608)0.027\\
        $b_9$ & 1.383(1.043,~1.724)~$<$0.00001 & 1.347(1.021,~1.673)~0.33 & 0.9(0.619,~1.182)0.69 & - & 0.861(0.25,~1.585)0.01\\
        $b_{10}$ & $-$0.72($-$1.079,~$-$0.361)~$<$0.00001 & $-$0.575($-$0.933,~$-$0.217)~0.72 & $-$0.436($-$0.89,~0.017)0.86 & - & $-$0.42($-$0.795,~$-$0.022)0.034\\
        $b_{11}$ & $-$0.557($-$0.93,~$-$0.182)~$<$0.00001 & $-$0.481($-$0.837,~$-$0.126)~0.64 & $-$0.451($-$0.856,~$-$0.047)0.82 & - & $-$0.453($-$0.795,~$-$0.11)0.0092\\
        $b_{12}$ & $-$0.224($-$0.764,~0.315)~0.15 & 0.147($-$0.287,~0.581)~0.89 & $-$0.015($-$0.392,~$-$0.361)0.99 & - & 0.119($-$0.198,~0.437)0.46\\
        $b_{13}$ & 1.1(0.6,~1.59)~$<$0.00001 & 0.7(0.373,~1.028)~0.29 & 0.835(0.542,~1.127)0.53 & - & 0.582(0.194,~0.953)0.0029\\
        $b_{14}$ & $-$10.995($-$218.166,~196.177)~0.69& $-$10.998($-$259,~237)~0.85 & $-$0.96($-$1.21,~$-$0.71)0.016 & - & $-$1.164($-$2.592,~$-$0.135)0.056\\
        %\headrow
        %Brier score & 0.021(0.0159~0.0436) & 0.0254(0.0162~0.0.0421) & 0.0257(0.0177~0.0394) & 0.0288(0.0194~0.0312) & 0.1266(0.0997~0.1783)\\
        \bottomrule
        AIC & 833.84 & 817.74 & 810.16 & 846.80 & 826.27\\
        p-value & $<$0.00001*** & $<$0.00001*** & $<$0.00001*** & $H_0$ & -\\
        \bottomrule  % Please only put a hline at the end of the table
    \end{tabular}}
    \label{table1}
\end{sidewaystable}

\clearpage

\bibliographystyle{abbrv}
%\bibliography{refs}

\end{document}